\documentclass[12pt]{article}
\usepackage{graphicx}

\def\boulder{Department of Physics\\ University of Colorado Boulder, UCB 390 , Boulder CO, 80309, USA}

\def\Title#1{\begin{center} {\Large #1 } \end{center}}
\def\Author#1{\begin{center}{ \sc #1} \end{center}}
\def\Address#1{\begin{center}{ \it #1} \end{center}}

\newenvironment{Abstract}{\begin{quotation}  }{\end{quotation}}
\newenvironment{Presented}{\begin{quotation} \begin{center} 
             PRESENTED AT\end{center}\bigskip 
      \begin{center}\begin{large}}{\end{large}\end{center} \end{quotation}}

\begin{document}
\begin{titlepage}

\vfill
\Title{Recent Results from Long-Baseline Neutrino Experiments}
\vfill
\Author{ Alysia D. Marino}
\Address{\boulder}
\vfill
\begin{Abstract}
 We are moving into an era of precision measurements of neutrino mixing, and it is increasingly necessary to use a 3-flavor framework to describe the results. This paper will focus on  recent results from long-baseline neutrino experiments, especially accelerator-based beams.  Using $\nu_{\mu}$ beams, we have now observed the disappearance of $\nu_{\mu}$, and the appearance of both $\nu_{e}$ and $\nu_{\tau}$.  It will also briefly describe some of the long-baseline experiments that have recently started operation, from which data are expected soon.

\end{Abstract}
\vfill
\begin{Presented}
XXXIV Physics in Collision Symposium \\
Bloomington, Indiana,  September 16--20, 2014
\end{Presented}
\vfill
\end{titlepage}
\def\thefootnote{\fnsymbol{footnote}}
\setcounter{footnote}{0}

\section{Neutrino Mixing}

Over the past 20 years, compelling evidence has emerged for neutrino flavor oscillations.    This evidence has been seen over a wide range of energies and distances and with both natural and manmade neutrino sources.  

We can interpret these results as arising from the fact that neutrinos are produced in flavor eigenstates of the weak interaction ($\nu_{e}$, $\nu_{\mu}$, and $\nu_{\tau}$), and these are a mixture of neutrino mass eigenstates ($\nu_{1}$, $\nu_{2}$, and $\nu_{3}$).  Interference between the flavor eigenstates causes the apparent neutrino flavor to oscillate with distance.  The mixing between the mass and flavor states can be well-described by a unitary mixing matrix, known as the MNSP matrix:

\begin{equation}
\label{eq:mnsp}
\left( \begin{array}{c}
\nu_{e}\\ 
\nu_{\mu} \\ 
\nu_{\tau}\\
\end{array} \right)
=
\left( \begin{array}{ccc}
U_{e1} & U_{e2} & U_{e3} \\
U_{\mu1} & U_{\mu2} & U_{\mu3} \\
U_{\tau1} & U_{\tau2} & U_{\tau3} \\
\end{array} \right)
\left( \begin{array}{c}
\nu_{1}\\ 
\nu_{2} \\ 
\nu_{3}\\
\end{array} \right).
\end{equation}
In general, this mixing matrix can be parameterized as the product of three rotation matrices and a complex phase:
\begin{center}
\begin{equation}
\label{eq:mnsp_angles}
U=
\left[ \begin{array}{ccc}
1 & 0 & 0\\
0&  c_{23} & s_{23} \\
0 & -s_{23} & c_{23}\\
\end{array}  \right] 
\times 
\left[ \begin{array}{ccc}
c_{13} & 0 & s_{13}e^{-i \delta_{CP}}\\
0&  1 & 0  \\
-s_{13}e^{i \delta_{CP}}& 0  & c_{23}\\
\end{array}  \right] 
\times 
\left[ \begin{array}{ccc}
c_{12} & s_{12} & 0\\
-s_{12}& c_{12} & 0\\
0 & 0 & 1\\
\end{array}  \right] 
\end{equation}
\end{center}
 where $c_{ij}$ and $s_{ij}$ are the cosines and sines of the three mixing angles $\theta_{12}$, $\theta_{13}$,  and $\theta_{23}$, and $\delta_{CP}$ is the phase.  Anti-neutrinos mix via the complex conjugate of this matrix.

The probability to observe a neutrino transition from one flavor ($\alpha$) to another ($\beta$) is given by~\cite{kayser}
\begin{equation}
\label{eq:oscprob}
\begin{array}{r}
P(\nu_{\alpha}\rightarrow\nu_{\beta})=\delta_{\alpha \beta}-4\sum\limits_{i>j}\Re (U_{\alpha i}^{*}U_{\beta i}U_{\alpha j} U_{\beta j}^{*})\sin^{2}[\Delta m_{ij}^{2} (L/4E)]   \\ 
+2\sum\limits_{i>j}\Im (U_{\alpha i}^{*}U_{\beta i}U_{\alpha j} U_{\beta j}^{*})\sin[\Delta m_{ij}^{2} (L/2E)] 
\end{array}
\end{equation}
where $L$ is the distance travelled, $E$ is the neutrino energy, and $\Delta m_{ij}^{2}$ is the difference in the square of the neutrino masses $m_i^2 - m_j^2$.  In this way, neutrino flavor change depends on the values of the three mixing angles, the differences between the three masses and the possible CP-violating phase $\delta_{CP}$, all of which must be determined experimentally.

The oscillations are a function of $\Delta m^{2}  L/E$.  For long-baseline neutrino beams, the oscillation effects are dominated by mixing between $\nu_{1}-\nu_{3}$ and $\nu_{2}-\nu_{3}$.  It is known that $\Delta m_{21}^{2}$ is considerably smaller than $\Delta m_{31}^{2}$, so for a $\nu_\mu$ beam the mixing between $\nu_{1}-\nu_{2}$ only becomes significant at very long distances or very low energies.  The sign of $\Delta m^{2}_{31}$ is currently unknown, thus we do not know if $\nu_{3}$ is the heaviest neutrino  (the so-called ``normal hierarchy") or the lightest neutrino (the ``inverted hierarchy") 

For a beam of muon neutrinos, the survival probability for $\nu_{\mu}$ is given approximately by
\begin{equation}
\label{eq:numu}
P(\nu_\mu \rightarrow \nu_\mu) \simeq 1 -\sin^2{2\theta_{23}}\cdot \sin^{2}{(\Delta m^{2}_{32} L/4E)},
\end{equation}
while  appearance of $\nu_{e}$ in vacuum is 
\begin{equation}
\label{eq:nuevac}
P(\nu_\mu \rightarrow \nu_e) \simeq \sin^2{2\theta_{13}}\cdot \sin^{2}{\theta_{23}} \cdot \sin^{2}{(\Delta m^{2}_{31} L/4E)}.
\end{equation}
Notice that the $\nu_{e}$ appearance depends on both the 1-3 and 2-3 sectors.  As neutrinos travel through matter, Equation~\ref{eq:nuevac} is modified slightly to~\cite{PDG}
\begin{equation}
\label{eq:nuematter}
P(\nu_\mu \rightarrow \nu_e) \simeq \sin^2{2\theta_{13}}\cdot \frac{\sin^{2}{\theta_{23}} }{(A-1)^{2}}\cdot \sin^{2}{((A-1) \Delta m^{2}_{31} L/4E)} 
\end{equation} where $A= \sqrt{2}G_{F} N_{e} \frac{2E}{\Delta m^{2}_{31}}$, which depends on the Fermi coupling constant, $G_{F}$, and $N_{e}$, the electron number density of the material. Note that this expression depends on the sign of $\Delta m^{2}_{31}$, and therefore on the neutrino mass hierarchy.  $A$ changes sign for anti-neutrinos, so even with a true  $\delta_{CP}=0$ the probability for $\nu_{\mu} \rightarrow \nu_{e}$ will be enhanced (suppressed) compared to $\bar{\nu}_{\mu} \rightarrow \bar{\nu}_{e}$ for the normal (inverted) hierarchy.  Including the effects of non-zero $\delta_{CP}$ adds additional terms to $P(\nu_\mu \rightarrow \nu_e)$ including ~\cite{PDG}
\begin{equation}
\label{eq:nuecp}
-\frac{|\Delta m^{2}_{21}|}{|\Delta m^{2}_{31}|} 
\frac{ \sin{\delta_{CP}} \sin{ 2 \theta_{12}} \sin{ 2 \theta_{13}} \sin{ 2 \theta_{23}} \cos{\theta_{13}}}{A(1-A)} \sin{\Delta}  \sin{(A \Delta)} \sin{((1-A)\Delta)},
\end{equation}
where $\Delta = \Delta {m^{2}_{31} L/4E}$.  This term also changes sign for anti-neutrinos, but has a different dependence on $L$ and $E$ compared to the matter term in Equation~\ref{eq:nuematter}.  So in this way, differences between  $\nu_{\mu} \rightarrow \nu_{e}$ and  $\bar{\nu}_{\mu} \rightarrow \bar{\nu}_{e}$ can help to resolve the hierarchy question and provide a measurement of CP-violation in the lepton sector if $\delta_{CP}$ is found to be non-zero.  Thus the interpretation of current and future neutrino beams will rely on using a 3-flavor description of neutrino oscillations and will focus on comparisons between neutrino and anti-neutrino signals.

\section{MINOS Results}

The  Main Injector Neutrino Oscillation Search (MINOS) experiment~\cite{minosnim} consists of two magnetized iron sampling calorimeters.  The near detector has a mass of 1 kton and is located on the Fermilab site, approximately 1 km from the target.  The far detector has a mass of 5.4 ktons and is 735 km away in the Soudan Mine in Minnesota.  MINOS is located on the Neutrinos at the Main Injector (NuMI) neutrino beam, which is initiated by 120 GeV protons.  Two magnetic horns are used to focus the charged pions and kaons that result from the interactions of these protons in a graphite target. By changing the direction of the horn current, the beam can be either predominately $\nu_{\mu}$ or $\bar{\nu}_{\mu}$.

MINOS collected data from May 2005 until May of 2012.  A total of $10.7\times 10^{20}$ protons on target were collected with the beam in neutrino mode, while $3.61\times 10^{20}$ protons on target were collected in anti-neutrino mode.  Additionally the far detector recorded 37.88 kton-years of atmospheric neutrino data.

\begin{figure}[htb]
\centering
\includegraphics[height=4in]{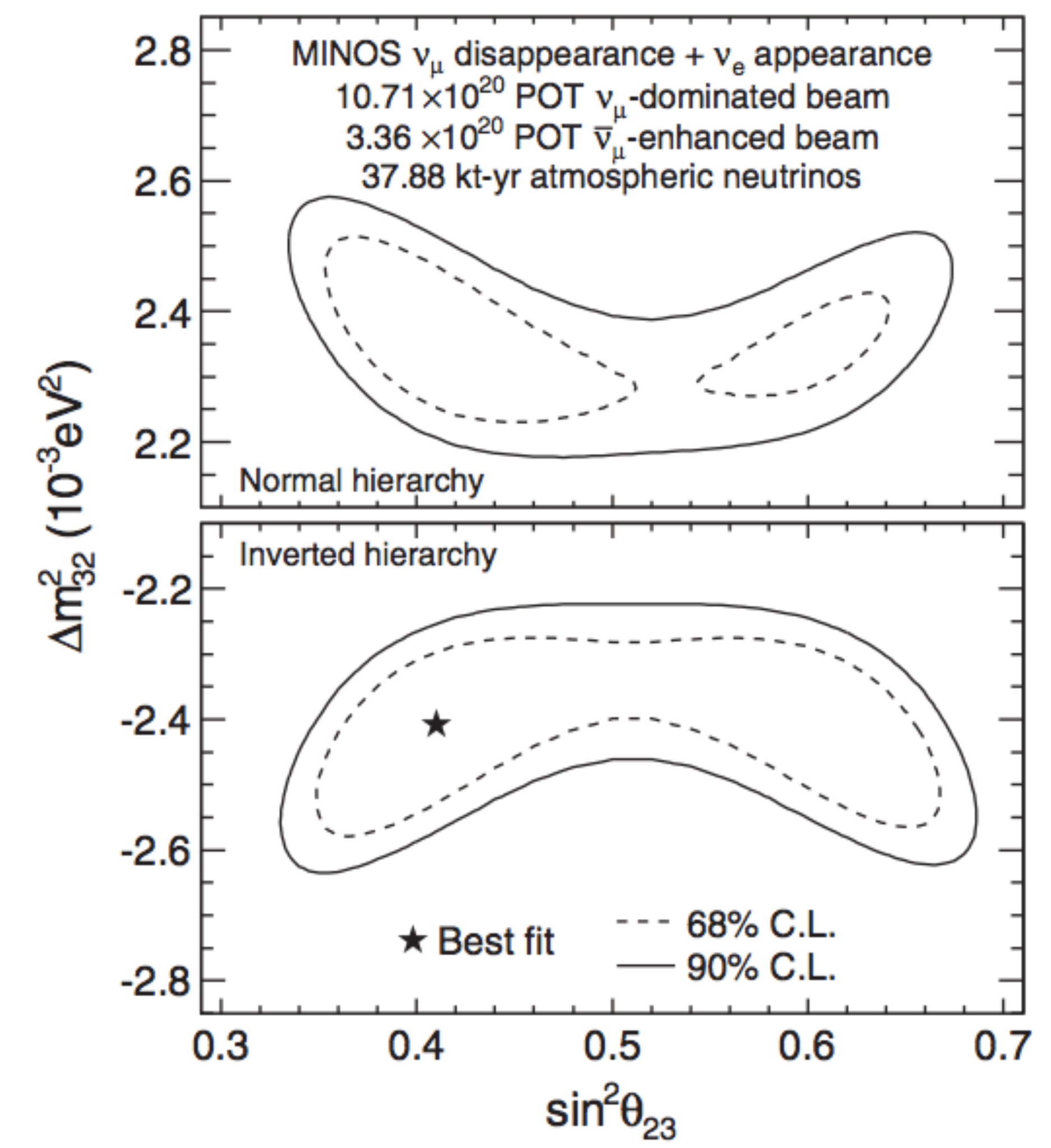}
\caption{The allowed parameter regions are shown here at the 90\% CL (solid) and 68\% CL (dashed) from a combined fit to all of the MINOS data.  Reproduced from~\cite{minoscombined}.}
\label{fig:minoscombined}
\end{figure}

MINOS recently published a combined 3-flavor fit to the beam neutrino and anti-neutrino data + atmospheric data~\cite{minoscombined}.  This included a fit to both the $\nu_{\mu}$ disappearance data and the $\nu_{e}$ appearance data.  The value of $\theta_{13}$ was constrained by results from reactor neutrino experiments.  The allowed regions are shown in Figure~\ref{fig:minoscombined}.

MINOS has also directly compared the $\nu$ and $\bar{\nu}$ signals for both $\nu_{\mu}$~\cite{minosantinumu} and $\nu_{e}$~\cite{minosantinue}.  For  $\nu_{\mu}$ disappearance, both have very similar best fit values and overlapping allowed regions.  For $\nu_{e}$ the data somewhat disfavors negative values of $\delta_{CP}$.

\section{T2K Results}
The Tokai-to-Kamioka (T2K) experiment~\cite{t2knim} uses a beam that originates at the Japan Proton Accelerator Research Complex (J-PARC) on the east coast of Japan.  T2K's $\nu_{\mu}$ beam is initiated by 30 GeV protons striking a graphite target.  Three magnetic horns focus the secondary hadrons.    Near detectors are located 280 m from the target  at J-PARC, while the Super-Kamiokande detector is used as the far detector, 295 km away.   The beam is aimed 2.5$^\circ$ away from Super-Kamiokande and the off-axis near detector (ND280).  This results in a beam with a narrower energy spectrum and enhanced flux at lower energies.

The ND280 near detector is magnetized.  Its tracker consists of 2 fine-grained scintillator trackers sandwiched between 3 TPCs.  A brass/lead/scintillator calorimeter forms a  pi-zero detector, which sits upstream of the tracker.  The detectors are surrounded by lead/scintillator electromagnetic calorimeters, and additional scintillator paddles are interspersed between the layers of the magnet yoke.  One of the fined-grained scintillator detectors and the pi-zero detector also contain passive water targets.  The far detector, Super-Kamiokande, is a water Cherenkov detector with a fiducial mass of 22.5 kton of water, surrounded by 11,000 phototubes.  A separate veto detector surrounds the main detector to reject external events.  As of October 2014, T2K has collected $6.88\times 10^{20}$ protons on target with the beam in neutrino mode.  With the constraints on the flux and cross section from the ND280 and external data, the uncertainties on the number of expected $\nu_{\mu}$ and $\nu_{e}$ events at SK are 7.7\% and 6.8\%, respectively.  

T2K sees significant disappearance of $\nu_{\mu}$~\cite{t2knumu} and that currently yields the most precise measurement of $\theta_{23}$.  T2K also sees 
 the appearance of $\nu_{e}$~\cite{t2knue} with a significance of 7.3$\sigma$ compared the the background expectations without $\nu_{\mu} \rightarrow \nu_{e}$.  
 
\begin{figure}[htb]
\centering
\includegraphics[width=4in]{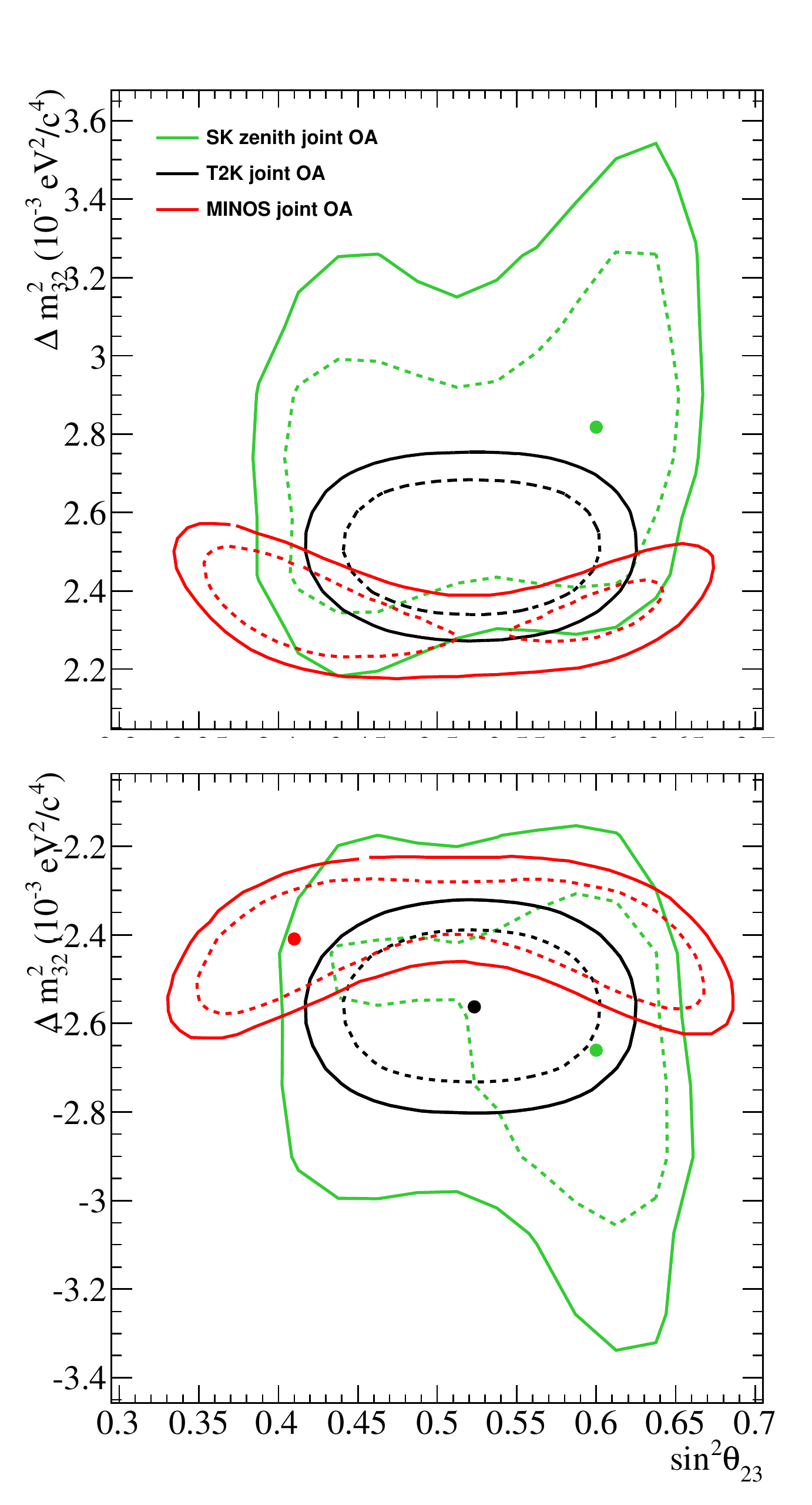}
\caption{This figure compares the allowed regions from T2K, MINOS, and Super-Kamiokande.  The top plot shows the fits for normal hierarchy, while the bottom plots shows fits for the inverted hierarchy. The solid (dashed) contours are the  90\% (68\%) CL regions.   Constraints from reactor experiments are not included here.}
\label{fig:23all}
\end{figure}

\begin{figure}[htb]
\centering
\includegraphics[width=5in]{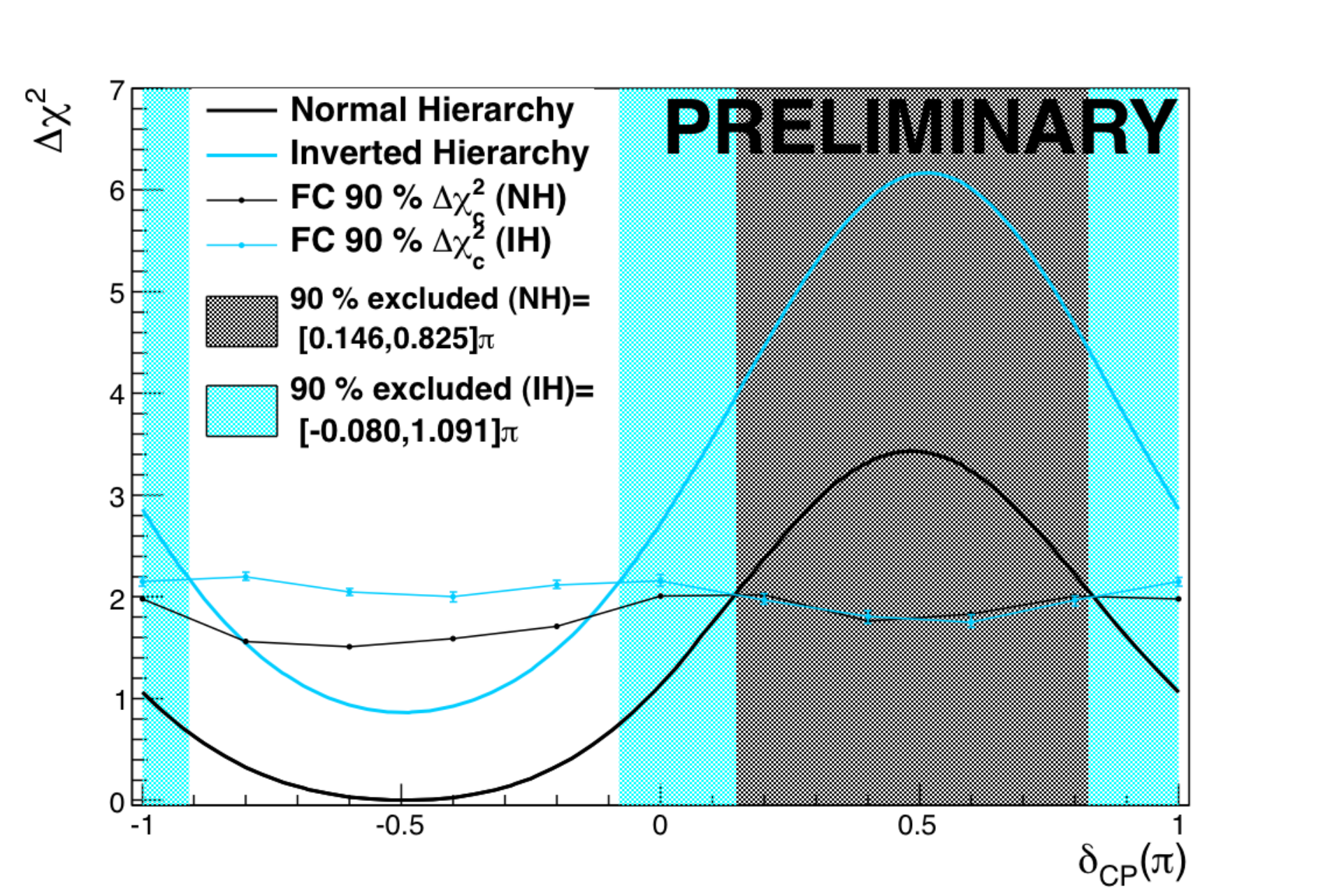}
\caption{This figure shows the $\Delta \chi ^{2}$ as a function of $\delta_{CP}$ for the T2K including constraints from reactor anti-neutrino disappearance measurements. }
\label{fig:deltacp}
\end{figure}

 T2K has also recently performed a combined 3-flavor fit to both its $\nu_{\mu}$ disappearance and $\nu_{e}$ appearance signals~\cite{t2k3nu} in neutrino mode.  Figure~\ref{fig:23all} compares the allowed regions for $\Delta m_{23}^{2}$ and $\sin^{2}{\theta_{23}}$ from the T2K beam data, MINOS atmospheric $\nu$ and beam data, and Super-Kamiokande atmospheric $\nu$ data.   The allowed regions from all three experiments agree very well.  T2K has also performed this combined appearance/disappearance 3-flavor fit including a constraint on $\theta_{13}$ from reactor anti-neutrino disappearance searches.  The $\Delta \chi ^{2}$ for these fits as a function of $\delta_{CP}$ is shown in Figure~\ref{fig:deltacp}.  The data  show a strong preference for negative values of  $\delta_{CP}$ at the 90\% CL.

In the summer of 2014 T2K also began collecting anti-neutrino data with  $0.51\times 10^{20}$ protons on target collected so far.  So far T2K has only collected $\simeq$10\% of its total proton on target goal, and future running will include a mixture of neutrino and anti-neutrino data. 

\section{Tau Appearance Results}

Searches have also been conducted for $\nu_{\mu} \rightarrow \nu_{\tau}$.  In~\cite{opera} the OPERA experiment reported the detection of 3 $\nu_{\tau}$ events in their 2008-2012 dataset.  At the Neutrino 2014 conference they also reported a 4th $\nu_{\tau}$ event so that the null oscillation hypothesis is now rejected at the 4.2$\sigma$ confidence level.  

Super-Kamiokande has also reported evidence for $\nu_{\tau}$ appearance at the 3.8 $\sigma$ confidence level in atmospheric neutrino interactions that contain hadronic $\tau$ decays~\cite{sktau}.

\section{Near Future Projects}

The MINOS experiment concluded in data taking 2012, however the detectors continue to operate in the NOvA-era NuMI beam as the MINOS+ experiment.  Compared to the MINOS era, this beam has a higher power, a greater energy, and a greater flux of neutrinos per proton on target.  The physics goals for MINOS+  include precision 3-flavor mixing tests, neutrino cross section measurements, and searches for sterile neutrinos and non-standard interactions.

The NOvA experiment uses the NuMI beam with highly-segmented active liquid scintillator detectors.  In addition to a near detector at Fermilab, NOvA has a 14 kton far detector,  located 810 km away in Ash River, Minnesota.  Both detectors are approximately 14 mrad off of the axis of the beam.  The first NOvA beam was delivered in 2013 and construction on both the near and far detectors was completed in 2014.  The first physics results are expected within the next year.

With its longer baseline, the matter effects are more significant in NOvA than they are in T2K.  For maximal $\theta_{23}$  mixing, NOvA can resolve the hierarchy at the 95\% confidence level sensitivity over 1/3 of the range of $\delta_{CP}$ values, and the combination of NOvA and future T2K running can resolve the hierarchy at a significance that is better than 68\% confidence over all possible values of $\delta_{CP}$~\cite{novasens}.

\begin{figure}[htb]
\centering
\includegraphics[width=6in]{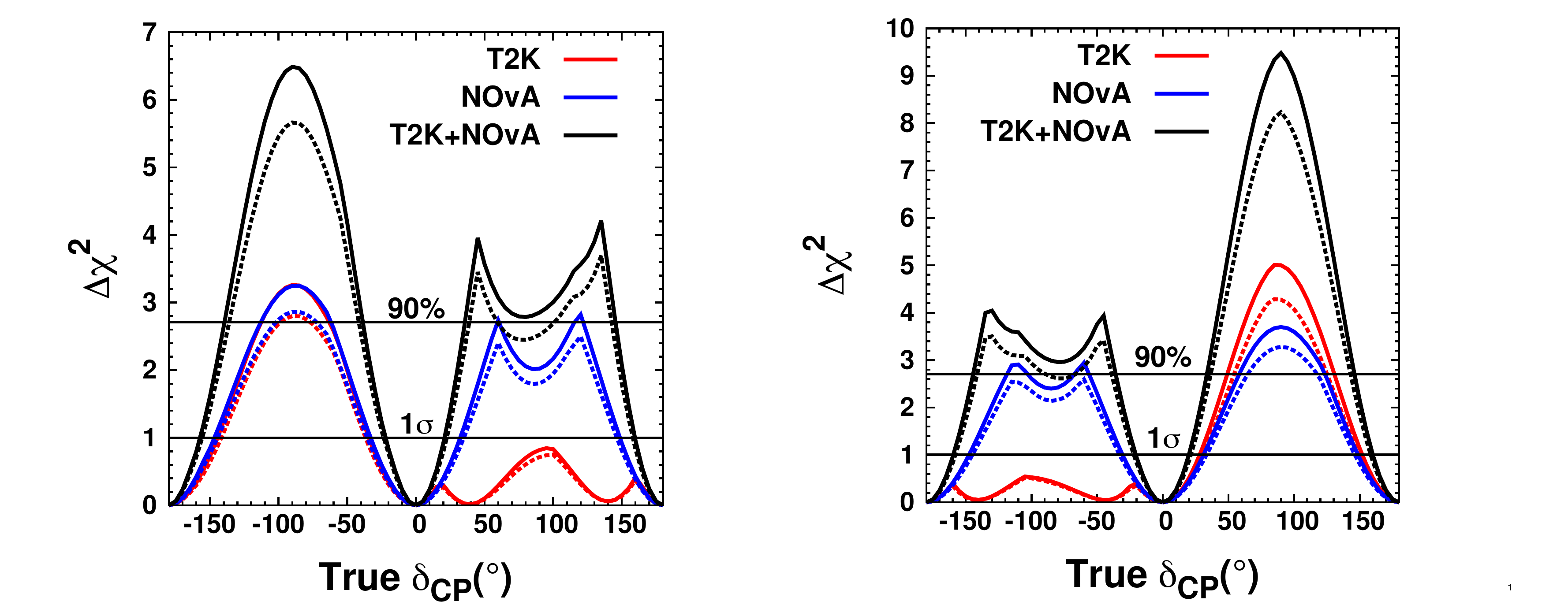}
\caption{This figure shows the $\Delta \chi ^{2}$ as a function of $\delta_{CP}$ for future T2K and NOvA running, assuming an equal mixture of neutrinos and anti-neutrinos.  The left (right) plot shows assumes a true normal (inverted) hierarchy.  This figure is reproduced from~\cite{t2ksens}.}
\label{fig:futuredeltacp}
\end{figure}

Future neutrino and anti-neutrino running with NOvA and T2K can constrain $\delta_{CP}$~\cite{t2ksens}. Figure~\ref{fig:futuredeltacp} shows the sensitivity to measuring non-zero $\delta_{CP}$ for T2K, NOvA, and the combined result.  The combination of T2K and NOvA has more than 1$\sigma$ sensitivity over 75\%  of the $\delta_{CP}$ range. 

\section{Conclusion}

We are moving into an era of precision measurements of neutrino oscillation parameters with accelerator-based neutrino experiments.  Increasingly it is necessary to use a 3-flavor framework to describe these oscillations.  

In recent years MINOS and T2K have observed the disappearance of $\nu_{\mu}$ and the appearance of $\nu_{e}$.  
MINOS operations concluded in 2012, and they have reported a joint 3-flavor analysis including both atmospheric neutrinos and beam neutrinos and anti-neutrinos.  T2K has presented results using a neutrino beam, and the comparison of the T2K results to the results from reactor anti-neutrinos currently favors large negative values of $\delta_{CP}$.  The appearance of $\nu_{\tau}$ has also been seen by the OPERA and Super-Kamiokande collaborations.

Our understanding of neutrino mixing will continue to improve over the next few years.  T2K began collecting anti-neutrino data in the summer of 2014 and initial results are expected soon.   MINOS+ and NOvA have started data collection and will also present their first physics results within the next year.  These future results will dramatically increase knowledge of the neutrino mass hierarchy and leptonic CP violation.

\end{document}